\documentclass[twocolumn,aps,prb,showpacs,amsfonts,amssymb,amsmath]{revtex4}

\usepackage{graphicx}
\begin{document}

\title{Scanning tunneling spectroscopy of a magnetic atom on graphene in the Kondo
regime}

\author{Huai-Bin Zhuang$^{1}$, Qing-feng Sun$^{1}$, and X. C. Xie$^{1,2}$ }
\affiliation{$^1$Beijing National Lab for Condensed Matter Physics
and Institute of Physics, Chinese Academy of Sciences, Beijing
100190, China\\
$^2$Department of Physics, Oklahoma State University, Stillwater,
Oklahoma 74078 }

\date{\today}

\begin{abstract}
The Kondo effect in the system consisting of a magnetic adatom on
the graphene is studied. By using the non-equilibrium Green function
method with the slave-boson mean field approximation, the local
density of state (LDOS) and the conductance are calculated. For a
doped graphene, the Kondo phase is present at all time.
Surprisingly, two kinds of Kondo regimes are revealed. But for the
undoped graphene, the Kondo phase only exists if the adatom's energy
level is beyond a critical value. The conductance is similar to the
LDOS, thus, the Kondo peak in the LDOS can be observed with the
scanning tunneling spectroscopy. In addition, in the presence of a
direct coupling between the STM tip and the graphene, the
conductance may be dramatically enhanced, depending on the coupling
site.
\end{abstract}

 \maketitle

Graphene, a single-layer hexagonal lattice of carbon atoms, has
recently attracted a great deal of attention due to its unique
properties and potential applications.\cite{ref1,ref2} The graphene
has an unique band structure with a linear dispersion near the
Dirac-points, such that its quasi-particles obey the two-dimensional
Dirac equation and have the relativistic-like behaviors. For a
neutral graphene, the Fermi level passes through the Dirac points
and the density of states (DOS) at the Fermi face vanishes, so that
its transport properties are greatly deviated from that of a normal
metal.

The Kondo effect has been paid great attention in condensed matter
community over the years.\cite{ref3,ref4,ref5} It is a prototypical
many-body correlation effect involving the interaction between a
localized spin and free electrons. The Kondo effect occurs in the
system of a magnetic impurity embedded in a
metal\cite{ref3,ref4,ref5} or a quantum dot coupled to the
leads,\cite{ref6,ref7,ref8} in which the magnetic impurity or the
quantum dot acts as a localized spin. At low temperature, the
localized spin is screened by the free electrons, resulting a spin
singlet and a very narrow Kondo peak located near the Fermi surface
in the local DOS (LDOS).

For a conventional metal, it has a finite DOS at the Fermi surface,
thus, at zero temperature the Kondo effect can occur with any weak
exchange interaction $J_{ex}$ between the magnetic impurity and the
free electrons. But in certain special systems, such as the nodal
d-wave superconductors or the Luttinger liquids, the DOS vanishes at
the Fermin surface with power law behavior. In these systems, there
exists a critical exchange coupling $J_c$, and the Konde effect only
occurs when the exchange interaction $J_{ex}>J_c$.\cite{ref9}

The graphene, which was successfully fabricated in recent years, has
an unique band structure.\cite{ref1} For the undoped neutral
graphene, its DOS is directly proportional to the energy
$|\epsilon|$ and the DOS vanishes at the Fermi surface, so it has an
unconventional Kondo effect. Very recently, some studies have
investigated the Kondo effect in the graphene and a finite critical
Kondo coupling strength was revealed.\cite{ref11,ref12} By varying
the gate voltage, the charge carriers of graphene can be easily
tuned experimentally. Then the Fermi level can be above or below the
Dirac points, and the DOS depends on the energy $\epsilon$ in the
form $|\epsilon-\epsilon_0|$, with the Dirac-point energy denoted by
$\epsilon_0$. In this situation, the DOS is quite small but finite
at the Fermi surface. In addition, there exists a zero point in DOS
that is very close to the Fermi surface. How is the Kondo effect
affected by this unique energy band structure?

\begin{figure}[h]
\includegraphics[width=6.0cm,height=4.0cm]{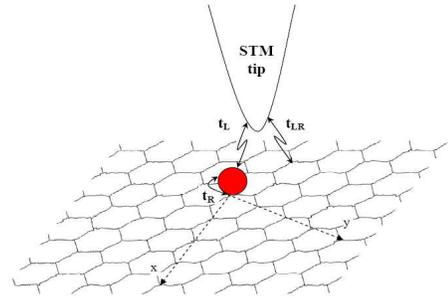}
\caption{ (Color online) The schematic diagram for a magnetic adatom
on the graphene probed by a STM tip. } \label{modelshow}
\end{figure}

In this Letter, we study the Kondo effect in the graphene. We
consider the model of a magnetic adatom on the graphene and to
investigate its scanning tunneling spectroscopy by attaching a STM
tip near the magnetic adatom (see Fig.1). Taking the slave-boson
(SB) mean field approximation and using the non-equilibrium Green
function method, the expressions of the LDOS and conductance are
obtained. Our results exhibit that for the doped graphene the Kondo
phase always exists, however, two kinds of Kondo regimes emerge. For
the undoped neutral graphene, the Kondo phase only exists when the
energy level $\epsilon_d$ of the adatom is larger than a critical
value $\epsilon_{dc}$.

The above mentioned system can be described by the Anderson
Hamiltonian. Here the magnetic atom is modeled by a single level
$\epsilon_d$ with the spin index $\sigma=\uparrow,\downarrow$ and a
constant on-site Coulomb interaction $U$. In the limit $U\rightarrow
\infty$, the double occupancy is forbidden. Then the SB
representation can be applied, and the Hamiltonian of the Anderson
model describing the system is transformed to the following
form:\cite{ref13}
\begin{eqnarray}
 H &=&\sum_{\sigma }\epsilon _{d}\hat{n}_{\sigma}+
 \sum_{i,\sigma}\epsilon_{0}a_{i\sigma }^{\dagger}a_{i\sigma }
 +t\sum_{\langle ij\rangle ,\sigma }a_{i\sigma }^{\dagger}a_{j\sigma }
 \notag \\
 &+&\sum_{k,\sigma }\epsilon _{k}c_{k\sigma }^{\dagger}c_{k\sigma
}+ \sum_{k,\sigma }\left( t_{L}c_{k\sigma }^{\dagger}f_{\sigma
}b^{\dagger} + t_{LR}c_{k\sigma }^{\dagger}a_{J\sigma }+H.c.\right)   \notag \\
&+& \sum_{\sigma }\left( t_{R}f_{\sigma }^{\dagger}ba_{0\sigma
}+H.c.\right)   +\lambda ( b^{\dagger}b+\sum_{\sigma }
\hat{n}_{\sigma} -1)
\end{eqnarray}
where $\hat{n}_{\sigma}= f_{\sigma }^{\dagger}f_{\sigma }$.
$a_{i\sigma}$ and $c_{k\sigma}$ are the annihilation operators at
the discrete site $i$ of the graphene and the tip of STM,
respectively. Here the graphene is assumed to be two-dimensional and
without disorder. In Hamiltonian (1), the graphene is described by
the tight-binding model and only the nearest neighbor hopping $t$ is
considered. $f_{\sigma}$ and $b$ are the pseudofermion operator and
SB operator of the magnetic atom. The last term represents the
single-occupancy constraint $b^{\dagger} b + \sum_{\sigma}
\hat{n}_{\sigma} =1$ with Lagrange multiplier $\lambda$. The
magnetic atom couples to the graphene at the lattice site $0$ with
the coupling coefficient $t_R$. The STM tip couples to both of the
magnetic atom and graphene, and $t_L$ and $t_{LR}$ are their
coupling coefficients. Here we consider that the STM tip only
contacts to a lattice site $J$ of the graphene. Following, we take
the standard SB mean field approximation in which the operator $b$
is replaced by the c-number $\langle b\rangle$. This approximation
is a qualitative correct for describing the Kondo regime at zero
temperature.\cite{ref13} For convenience, we also introduce the
renormalized energy $\tilde{\epsilon_d}\equiv\epsilon_d+\lambda$ and
hopping elements $\tilde{t}_{L}\equiv t_{L}\left\langle
b\right\rangle$ and $\tilde{t}_{R}\equiv t_{R}\left\langle
b\right\rangle $.

Next, we consider that a bias $V$ is applied between the STM tip and
the graphene to obtain the current $I$ flowing from STM to the
graphene and the LDOS around the magnetic atom. By using the
non-equilibrium Green function method, the current $I$ and LDOS
are:\cite{ref15}
\begin{eqnarray}
&&I=\frac{4e}{h}\int d\omega \mathbf{Re}\left[
\tilde{t}_{L}G_{fL}^{<}(\omega ) +t_{LR}G_{JL}^{<}( \omega )
\right], \\
&&\text{LDOS}( \omega ) = -(1/\pi)\mathbf{Im} G_{ff}^{r}(\omega),
\end{eqnarray}
where ${\bf G}^{<}$ and ${\bf G}^r$ are the standard lesser and
retarded Green's functions, and $G_{ff/fL/JL}^{</r}$ in Eq.(2) and
(3) are the elements of the matrix ${\bf G}^{</r}$. Under the SB
mean field approximation, ${\bf G}^{<}$ and ${\bf G}^r$ can be
solved from the Keldysh and Dyson equations: ${\bf G}^{<}=(1+{\bf
G}^{r}{\bf \Sigma}^{r}){\bf g}^{<}(1+{\bf \Sigma}^{a}{\bf G}^{a})$
and ${\bf G}^r =({\bf g}^r-{\bf \Sigma}^r)^{-1}$, where ${\bf
\Sigma}^{r,a}$ are the self-energies. $\Sigma^{r}_{Lf}=
\Sigma^{r}_{fL}=\tilde{t}_L$, $\Sigma^{r}_{f0}=
\Sigma^{r}_{0f}=\tilde{t}_R$, $\Sigma^{r}_{LJ}=
\Sigma^{r}_{JL}=t_{LR}$, and other elements of ${\bf \Sigma}^{r}$
are zero. ${\bf g}^{r/<}$ are the Green's functions of the isolated
STM tip, magnetic atom, and graphene (i.e. while $\tilde{t}_L
=\tilde{t}_R =t_{LR}=0$). They can easily be obtained:
$g_{LL}^r(\omega) =\sum_{k}g_{k}^{r}=-i\pi \rho _{L}$,
$g_{ff}^{r}(\omega)=1/(\omega -\tilde{\epsilon}_{d}+i0^{+})$,
$g_{LL}^<(\omega)=\sum_{k}g_{k}^{<}=2i\pi \rho _{L}f_{L}(\omega)$,
${\bf g}_{{\bf ij}}^{<}(\omega)=-f_{R}(\omega)( {\bf g}_{\bf
ij}^{r}-{\bf g}_{\bf ij}^{a})$, and
\begin{eqnarray}
&&{\bf g}_{{\bf ij}}^r(\omega)=\int\int dk_x dk_y e^{ia[(i_x-j_x)k_x
+(i_y-j_y)k_y]} {\bf g}^r_{k_x k_y}(\omega), \\
&& {\bf g}^r_{k_x
k_y}(\omega)=\frac{1}{(\omega-\epsilon_0)^2-|\phi|^2}
 \left(\begin{array}{ll} \omega-\epsilon_0 & \phi \\ \phi^*
 &\omega-\epsilon_0
 \end{array}\right),
\end{eqnarray}
where $\rho _{L}$ is the DOS of the STM tip. $f_{L}(\omega)$ and
$f_R(\omega)$ are the Fermi-Dirac distribution functions in the
isolated STM tip and graphene, respectively. ${\bf g}_{{\bf
ij}}^{r,<}(\omega)$ are the Green's functions of the isolated
graphene, and ${\bf i}=(i_x,i_y)$ and ${\bf j}=(j_x,j_y)$ are the
indices of the unit cell. Since there are two carbon atoms per unit
cell, ${\bf g}_{{\bf ij}}^r$ is a $2\times 2$ matrix. In Eq.(4) and
(5), $a$ is the lattice constant of the graphene, the integral $dk_x
dk_y$ runs over the first Brillouin zone, and $\phi= t (
1+e^{-iak_{x}}+e^{-iak_{x}+iak_{y}})$.

At end we need to self-consistently calculate two unknowns $\langle
b\rangle$ and $\lambda$ with the self-consistent equations:
\begin{eqnarray}
&& \langle b\rangle^2 +n_{\uparrow} +n_{\downarrow} =1 \\
&& 2 Im \int \frac{d\omega}{2\pi} \left(\tilde{t}_L G^<_{Lf}(\omega)
 +\tilde{t}_RG^<_{0f}(\omega) \right) +\lambda  \langle b\rangle^2
 =0,
\end{eqnarray}
where $n_{\sigma} = \int (d\omega/2\pi) Im G^<_{ff}(\omega)$ is the
electron occupation number in the spin state $\sigma$ of the
magnetic atom. In the numerical calculations, we take the hopping
energy $t=1$ as the energy unit. The chemical potential $\mu_R$ of
the graphene is set to be zero as the energy zero point, then the
bias voltage $V=\mu_L$. The temperature $T$ is assumed to be zero
since $k_B T$ is much smaller than $t=2.75eV$ in the real
experiments.

First, we assume the system to be decoupled to the STM tip (i.e. at
$t_L=t_{LR}=0$) and focus on the Kondo effect. Fig.2 shows
magnetic-atom LDOS for different $\epsilon_d$ and $\epsilon_0$. Here
the Dirac-point energy $\epsilon_0$ can be experimentally tuned by
the gate voltage. For the magnetic atom, the level $\epsilon_d$ is
fixed. But for an artificial atom, such as a quantum dot,
$\epsilon_d$ is also tunable.\cite{ref8} As shown in Fig.2a, there
exists three peaks in the LDOS. Two peaks locate at $\omega=\pm t$,
due to the peaks in the DOS of the graphene. The third peak is the
Kondo peak, which is always located around $\mu_R=0$ regardless of
the parameters of $\epsilon_d$, $\epsilon_0$, and $t_R$. Fig.2b
magnifies the curves of Fig.2a in the vicinity of $\omega=0$. It
clearly shows that the Kondo peak is much higher than the other two
peaks at $\omega=\pm t$. The lower the level $\epsilon _{d}$ is, the
higher and sharper is the Kondo peak and the closer it is to $\mu
_{R}=0$. In particular, for the undoped neutral graphene with
$\epsilon_0=0$, the Kondo peak is greatly deflected from $0$ (see
Fig.2b) although it seems to be near $0$ on a large scale (see
Fig.2a). The LDOS is zero at $\omega=0$ since the graphene's DOS
vanishes at this point. This characteristics is much different from
the conventional Kondo effect, in which the LDOS is quite large at
$\omega=0$.

\begin{figure}[h]
\includegraphics[width=8.0cm,height=6.5cm]{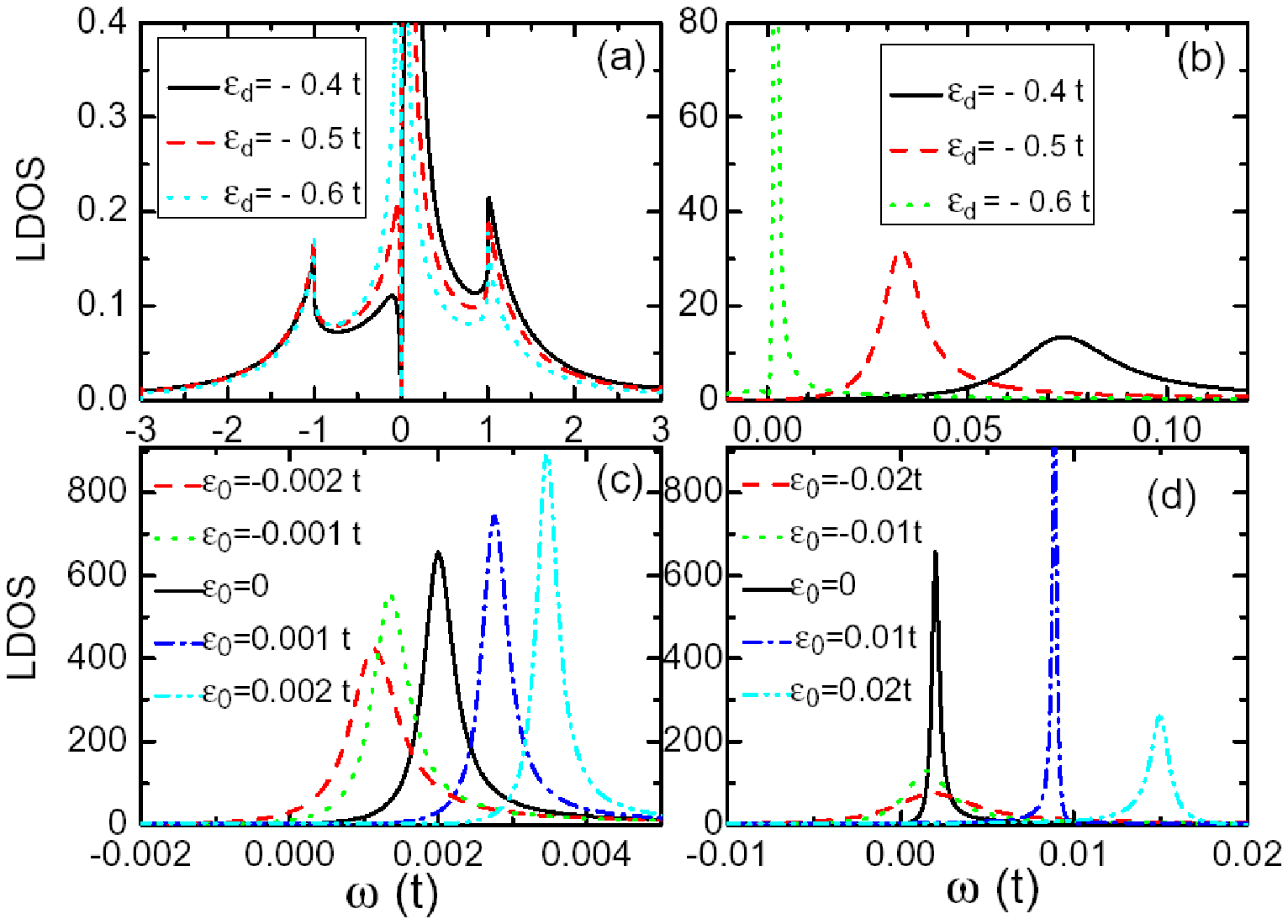}
\caption{ (Color online) LDOS vs. the energy $\omega$ with
$t_L=t_{LR}=0$ and $t_R=0.9t$. In (a), the Dirac-point energy
$\epsilon_0=0$. (b) is the enlarged view of the Kondo peaks in (a).
(c) and (d) are at different $\epsilon_{0}$ for the level
$\epsilon_{d}=-0.6t$.} \label{ldos}
\end{figure}

Next, the energy $\epsilon_0$ is tuned to be away from zero. While
$\epsilon_0<0$, the carriers in graphene are electron-like, but they
are hole-like for $\epsilon_0>0$. With $\epsilon_0$ decreasing from
$0$, the Kondo peak is gradually getting closer to $0$ and is
slightly widened (see Fig.2c). It recovers the shape of conventional
Kondo peak at large negative $\epsilon_0$ (e.g. $\epsilon_0=-0.01t$
or $-0.02t$ in Fig.2d). On the other hand, with $\epsilon_0$
increasing from $0$, the Kondo peak departs gradually from $0$ and
its shape never recovers to that of the conventional Kondo peak even
at large $\epsilon_0$.

From the position $\omega_1$ and half-width $\tilde{\Gamma}_{R}$ of
the Kondo peak, we can estimate the Kondo temperature by\cite{ref4}
$ T_{k}=\sqrt{\omega _{1}^{2}+\tilde{\Gamma}_{R}^{2}\left( \omega
_{1}\right) }$, where the peak position $\omega_{1}$ is obtained
from the equation $\omega -\tilde{\epsilon}_{d}- \textbf{Re} [
\tilde{ t}_{R}^{2} g_{00}^{r} ( \omega ) ]=0$ and the peak
half-width $\tilde{\Gamma}_{R} = -2 \tilde{t}_R^2 {\bf
Im}[g^r_{00}(\omega)]$. Fig.3a shows the Kondo temperature $T_{k}$
versus the level $\epsilon_{d}$ at a fixed $t_{R}$ for different
energy $\epsilon_0$. $T_{k}$ possesses the following
characteristics: (i) $T_{k}$ decreases monotonically with reducing
of $\epsilon_d$. When $\epsilon_d$ is approaching $-\infty$, $T_k$
goes to zero. (ii) For $\epsilon_0\not=0$, there exists two
different Kondo regimes. When $\epsilon_d$ is larger than a critical
value $\epsilon_{dc}$, $T_k$ is almost linearly dependent on
$\epsilon_d-\epsilon_{dc}$, so that $T_k$ drops rapidly while
$\epsilon_d$ is near $\epsilon_{dc}$ (see Fig.3a). This
characteristics is very different comparing to the conventional
Kondo effect, and it origins from the linear DOS of the graphene and
vanish-ness of the DOS at the Dirac point. On the other hand, when
$\epsilon_d <\epsilon_{dc}$, $T_k \propto e^{\epsilon_d}$ (i.e. $\ln
T_k \propto \epsilon_d$ as shown in Fig.3a). In this case the
system's behavior is similar to that of the conventional Kondo
effect. Since $k_B T_k \ll |\epsilon_0-\mu_R|$ in this regime, the
DOS at the Fermi surface is approximatively constant on the energy
scale $k_B T_k$. So it recovers the conventional Kondo behavior.
(iii) The behavior of $T_k$ for $\epsilon_0<0$ (i.e. the electron
doped graphene) and $\epsilon_0>0$ (i.e. the hole doped graphene) is
similar.

\begin{figure}[h]
\includegraphics[width=8.0cm,height=6.5cm]{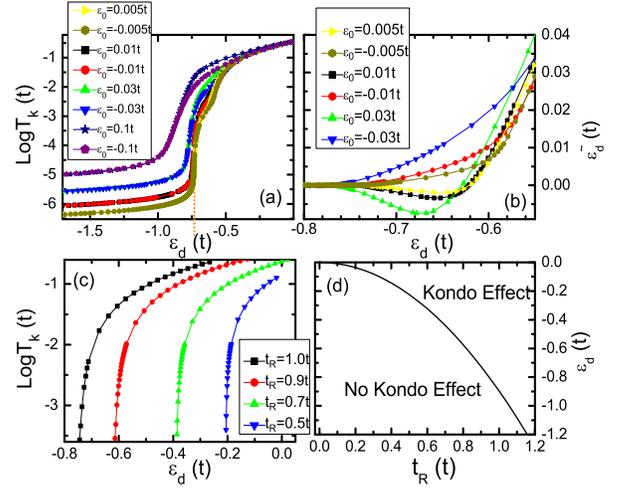}
\caption{( Color online) (a) $T_{k}$ vs. the level $\epsilon_d$ and
(b) $\tilde{\epsilon}_d$ vs. $\epsilon_d$ at $t_R=0.9t$ for
different energy $\epsilon_0$. (c) $T_{k}$ as functions of
$\epsilon_{d}$ for $\epsilon_{0}=0$ at different coupling $t_{R}$.
(d) Phase diagram of the $\epsilon_0=0$ in the parameter space of
($\epsilon_d$, $t_R$). } \label{Tk0}
\end{figure}

Fig.3b shows the renormalized level $\tilde{\epsilon}_d$ which
determines the position of the Kondo peak. While $\epsilon_0<0$, the
renormalized level $\tilde{\epsilon}_d$ is always positive and it
decreases monotonically with reducing of $\epsilon_d$.
$\tilde{\epsilon}_d$ approaches zero as $\epsilon_d \rightarrow
-\infty$. On the other hand, while $\epsilon_0>0$,
$\tilde{\epsilon}_d$ can be negative. The curve of
$\tilde{\epsilon}_d$-$\epsilon_d$ has a negative minimum. But
$\tilde{\epsilon}_d\rightarrow 0 $ is still true as $\epsilon_d
\rightarrow -\infty$.

Now we focus on the case of $\epsilon_0 = 0$. When the Dirac-point
energy $\epsilon_0$ approaches to $0$, the zero-DOS point approaches
to the Fermi surface, then both of the Kondo temperature $T_k$ and
the renormalized level $\tilde{\epsilon}_d$ goto zero in the whole
region of $\epsilon_d<\epsilon_{dc}$. At $\epsilon_0 = 0$ (i.e. for
the undoped neutral graphene), $T_k$ and $\tilde{\epsilon}_d$ reach
zero at $\epsilon_d=\epsilon_{dc}$. On the $\epsilon_d
<\epsilon_{dc}$ side, the self-consistent equations (6) and (7) do
not have a solution. In other words, the system cannot be in the
Kondo phase for $\epsilon_d <\epsilon_{dc}$. On the other side with
$\epsilon_d >\epsilon_{dc}$, the Kondo phase does exist. The Kondo
temperature $T_k$ is roughly proportional to
$\epsilon_d-\epsilon_{dc}$ (see Fig.3c). The critical value
$\epsilon_{dc}$ depends on the coupling strength $t_R$ between the
magnetic atom and the graphene. The larger $t_R$ is, the larger
$|\epsilon_{dc}|$ is (see Fig.3c). From $T_k=0$ or
$\tilde{\epsilon_d}=0$ at $\epsilon_d=\epsilon_{dc}$ and with the
aid of Eq.(7), the critical value $\epsilon_{dc}$ can be
analytically obtained:
\begin{equation}
\epsilon _{dc}=-\frac{2}{\pi }t_{R}^{2}
 \int_{-\infty}^{0}d\omega
\frac{\text{\textbf{I}m}\left( g_{II}^{r}\left( \omega \right)
\right) }{\omega }
\end{equation}
The above integral over $\omega$ is about 1.408, hence $\epsilon
_{dc} \approx 0.896 t_{R}^{2}$. Fig.3d shows the phase diagram in
the parameter space of ($\epsilon_d$, $t_R$). The curve gives the
critical value $\epsilon_{dc}$ versus the coupling strength $t_R$.
In the top right region, the Kondo phase emerges at low temperature.
But in the bottom left region, the system cannot enter into the
Kondo state even at zero temperature.

\begin{figure}[h]
\includegraphics[width=8.0cm,height=6.5cm]{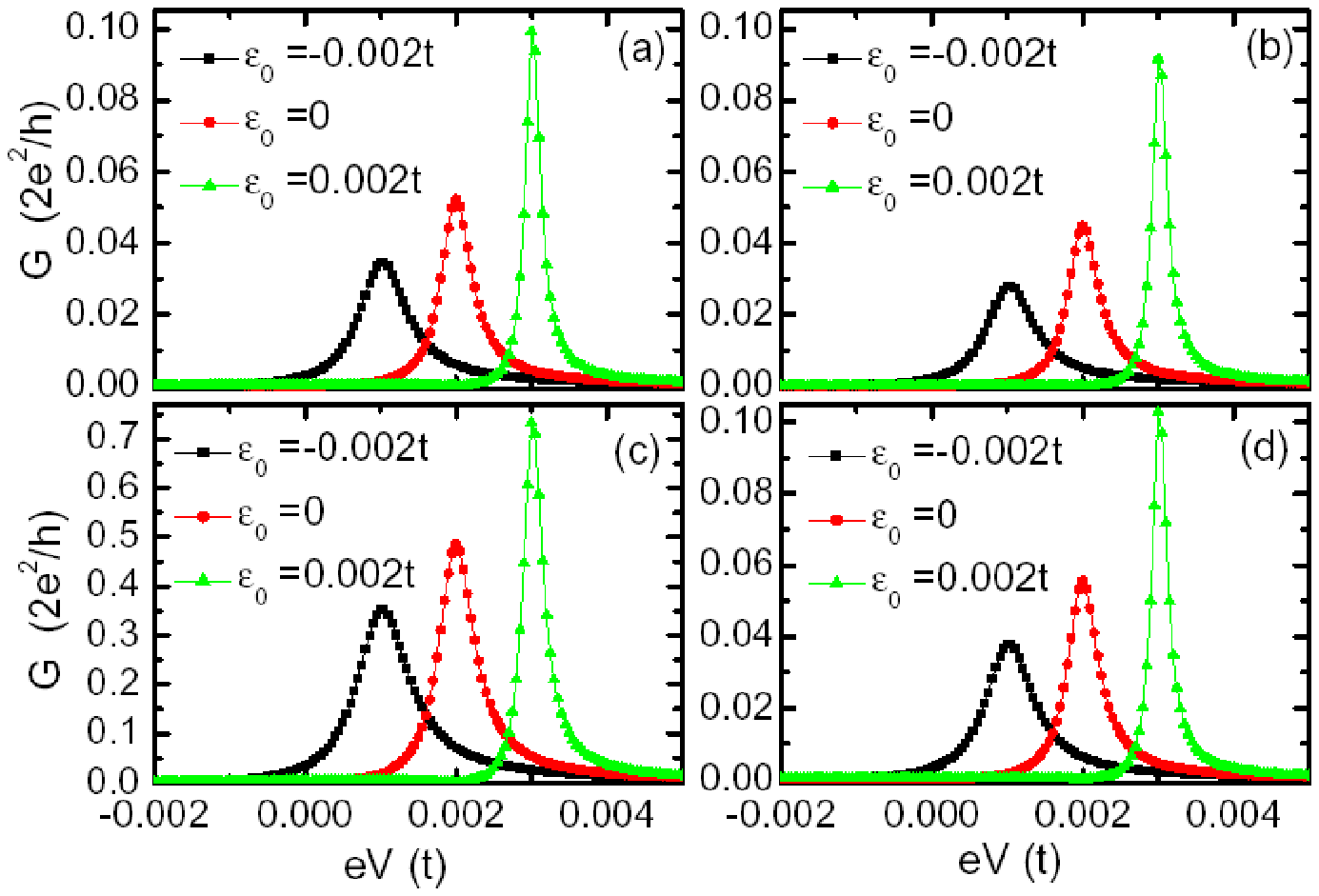}
\caption{ (Color online) Conductance $G$ vs. the bias $V$ for
different energy $\epsilon_{0}$ with $t_{R}=0.9t$, $t_{L}=0.002t$,
and $\epsilon_{d}=-0.6t$. In (a), $t_{LR}=0$, and in (b), (c), and
(d), $t_{LR}=0.03t$. The contact point $J$ of the STM tip and the
graphene in (b), (c), and (b) are 0, 1, and 2, respectively. }
\label{conductance}
\end{figure}

Finally, we investigate the conductance $G$ ($G\equiv dI/dV$) while
the STM tip is coupled. The coupling between the STM tip and the
adatom is normally much weak compared to that between the graphene
to the adatom, so we set $t_L$ and $t_{LR}$ on the order of $0.01t$.
Fig.4 shows the conductance $G$ versus the bias $V$ in the Kondo
regime. While without the direct coupling ($t_{LR}=0$), the curve of
$G$-$V$ is very similar to the curve of LDOS-$\omega$ (see Fig.4a
and 2c) since for the weak-coupled STM tip, the conductance $G$ is
proportional to the LDOS of the adatom.\cite{ref16} When the direct
coupling is present ($t_{LR}\not=0$), the coherence between the
direct path and the path passing the magnetic adatom occurs, which
usually leads to the Fano resonance.\cite{ref17} But in the present
graphene system, the Fano resonance does not appear anywhere, the
reason is that the transmission coefficient of the direct path also
depends on the energy of the incident electron. In Fig.4b, 4c, and
4d, $J$ of the contact point of the STM tip to the graphene is $0$
(same contact point with the adatom), $1$ (the nearest neighbor
site), and $2$ (the next nearest neighbor site), respectively. For
the same and the next nearest neighbor contact points, the
conductance $G$ is almost unaffected by the opening of the direction
path, although $t_{LR}=0.03t$ is much larger than $t_{L}=0.002t$.
Because that in the Kondo regime, a high Kondo peak emerges in the
LDOS of the adatom, so the path passing the adatom is large and
plays the dominant role. However, when the contact point is at the
nearest neighbor site, the conductance is greatly enhanced by the
opening of the direct path (see Fig.4c, note the different scale
comparing with the other plots in Fig.4) since the Kondo singlet
state is quite extensive. For the graphene system, the sharp Kondo
peak not only appears in the LDOS of the adatom, but also in the
LDOS of all odd-neighboring sites. The large LDOS in the
odd-neighboring sites and $t_{LR}>t_L$ lead that the direct path is
dominant, and the conductance $G$ is enhanced.

In summary, the Kondo effect in the system of a magnetic adatom on
the graphene and its scanning tunneling spectroscopy are
theoretically studied. For a doped graphene, the Kondo phase always
exists at zero temperature regardless of $n$-type or $p$-type
doping, and it exhibits two Kondo regimes: the conventional and
unconventional. For the undoped neutral graphene, the Kondo phase
only exists while the adatom's energy level is above a critical
value. These Kondo characteristics can be observed from scanning
tunneling spectroscopy since the conductance versus the bias is
similar to the LDOS of the adatom. In addition, if a direct coupling
between the STM tip and the graphene is present, the conductance may
be greatly enhanced, depending on the coupling site.

{\bf Acknowledgments:} We gratefully acknowledge the financial
support from NSF-China under Grants Nos. 10525418 and 10734110,
National Basic Research Program of China (973 Program project No.
2009CB929103), and US-DOE under Grants No. DE-FG02- 04ER46124.

\end{document}